\documentclass[prl, twocolumn]{revtex4}

\usepackage{amsmath,amssymb}  
\usepackage{psfrag}  
\usepackage{color}  
\usepackage[dvips]{epsfig}  
\usepackage{epsfig}

\usepackage{theorem}  

\theorembodyfont{\rmfamily}  
  
\theoremstyle{break}  
  
\def\QED{~\rule[-1pt]{5pt}{5pt}\par\medskip}

\def\su{\text{su}}  
\def\SU{\text{SU}} 
 
\begin{document}   

\title{Minimum construction of two-qubit quantum operations}

\author{Jun Zhang$^{1, 2} $, Jiri Vala$^{2}$,  Shankar Sastry$^1$
and K. Birgitta Whaley$^{2}$} 
  
\affiliation{$^1$Department of Electrical Engineering  
and Computer Sciences, University of California, Berkeley, CA 94720\\  
$^2$Department of Chemistry and Pitzer Center for Theoretical Chemistry, 
University of California, Berkeley, CA 94720} 
 
\date{\today}  
  
\begin{abstract}  
  Optimal construction of quantum operations is a fundamental problem
  in the realization of quantum computation. We here introduce a newly
  discovered quantum gate, B, that can implement any arbitrary
  two-qubit quantum operation with minimal number of both two- and
  single-qubit gates. We show this by giving an analytic circuit that
  implements a generic nonlocal two-qubit operation from just two
  applications of the B gate.  Realization of the B gate is
  illustrated with an example of charge-coupled superconducting qubits
  for which the B gate is seen to be generated in shorter time than
  the CNOT gate.
\end{abstract}   
  
\maketitle

Quantum computation requires achieving unitary operations on arrays of
coupled qubits in order to realize the speed-up associated with
quantum algorithms. It is usually described in the quantum circuit
model with combinations of single- and two-qubit
operations~\cite{Barenco:95}. While the algorithmic complexity is
independent of the efficiency of these circuits, which are known to be
interchangeable with a polynomial overhead~\cite{Kitaev:02}, the
performance of any physical realization of a quantum circuit may be
highly dependent on minimal switchings of control fields and
interaction Hamiltonians and achieving minimal time of gate
operations, due to the introduction of decoherence arising from
unwanted interactions between qubits and/or with the external
environment. Efficient construction of any arbitrary two-qubit quantum
operation is thus of high priority in the search for a realizable
quantum information processor.  The current standard paradigm is based
on a combination of quantum Controlled-NOT (CNOT) gates between pairs
of qubits and single-qubit gates~\cite{Barenco:95}.  Recent work shows
that the CNOT gate is also one of the most efficient quantum gates
known, in that just three applications supplemented with local gates
can implement any arbitrary two-qubit operation~\cite{Vidal:03,
  Shende:03a, Vatan:03, Zhang:03b}. In this paper, we introduce a
minimum construction from a newly discovered quantum gate, B, to
implement any arbitrary two-qubit quantum operation with two
applications of B together with six single-qubit gates, both of which
are the least possible.
  
\begin{figure}[tb] 
 \begin{center} 
  \psfrag{A1}[][]{$A_1$} 
  \psfrag{A2}[][]{$A_2$} 
  \psfrag{A3}[][]{$A_3$} 
  \psfrag{O}[][]{$O$} 
  \psfrag{L}[][]{$L$} 
  \psfrag{B}[][]{$B$} 
   \psfrag{c1}[][]{$c_1$} 
   \psfrag{c2}[][]{$c_2$} 
  \psfrag{c3}[][]{$c_3$} 
\includegraphics[width=50mm]{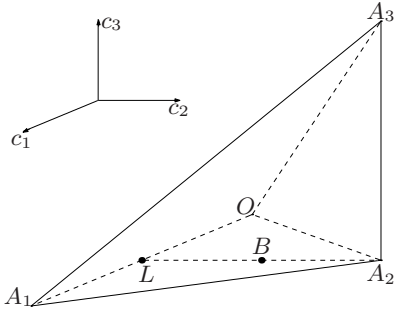}  
 \end{center}
 \caption{Tetrahedron $OA_1A_2A_3$ contains all the local equivalence
   classes of nonlocal gates~\cite{Zhang:02}, where $O([0, 0, 0])$ and
   $A_1([\pi, 0, 0])$ both correspond to local gates, $L([\frac\pi{2},
   0, 0])$ to the CNOT gate, $A_2([\frac\pi{2}, \frac\pi{2}, 0])$ to
   the Double-CNOT gate, $A_3([\frac\pi{2}, \frac\pi{2},
   \frac\pi{2}])$ to the SWAP gate, and $B([\frac\pi{2}, \frac\pi{4},
   0])$ to the new gate we introduce in this paper.  From the Cartan
   decomposition on $\su(4)$, any two-qubit unitary operation $U\in
   \SU(4)$ can be written as $U=k_1Ak_2=k_1\cdot e^{
     c_1\frac{i}{2}\sigma_x^1\sigma_x^2} \cdot
   e^{c_2\frac{i}{2}\sigma_y^1\sigma_y^2} \cdot
   e^{c_3\frac{i}{2}\sigma_z^1\sigma_z^2}\cdot k_2$, where
   $\sigma_{\alpha}^1\sigma_{\alpha}^2 = \sigma_{\alpha}\otimes
   \sigma_{\alpha}$, $\sigma_\alpha$ are the Pauli matrices, and
   $k_1$, $k_2\in {\SU(2)\otimes \SU(2)}$ are local
   gates~\cite{Kraus:01,Khaneja:01,Zhang:02}.  Since the local gates
   are fully accessible, it is evident that we need to construct a
   circuit for only the nonlocal block $A$ in terms of the available
   entangling gates (or Hamiltonians).}
\label{fig:N1}
 \end{figure}

The single-qubit operations are generally easily implemented by a
local Hamiltonian or external field and can be finitely generated by
any convenient basis on $\su(2)$~\cite{Low72}. In contrast, two-qubit
operations are highly dependent on the physical implementation and it
is in general much more difficult to implement an {\em arbitrary}
two-qubit operation. The number of single-qubit gates can be
independently optimized from a carefully chosen
library~\cite{Shende:03a} once the intrinsic quantum circuit structure
of two-qubit gates is determined. It is therefore advantageous to
employ the concept of \emph{locally equivalent} two-qubit gates,
namely, $U= k_1 U_1 k_2$, where $k_1$, $k_2$ are local unitary gates,
in order to obtain the minimal total circuit length. We denote the
local equivalence relation by $U\sim U_1$.  It has been shown that two
gates are locally equivalent if and only if they have identical values
of three invariants~\cite{Makhlin:00}.  Classification of two-qubit
gates based on these invariants was given an intuitive geometric
interpretation in~\cite{Zhang:02}, which is summarized in
Fig.~\ref{fig:N1}. Each point in the tetrahedron (also known as the
Weyl chamber, from the specific group symmetries used in its
construction) represents a local equivalence class of nonlocal gate
$U$, with the exception of vertices $O$ and $A_1$ which are local
gates.  The special status of CNOT was demonstrated by an analysis of
circuit optimality within the geometric approach, which showed that it
is indeed the most efficient Controlled-Unitary operation and requires
only three applications to construct any arbitrary two-qubit quantum
operation~\cite{Zhang:03b}.  These and
related~\cite{Vidal:03,Shende:03a,Vatan:03} investigations of
optimality with CNOT motivate the enquiry as to the existence of
two-qubit gates that may be even more efficient than CNOT.  The
demonstration of at least one gate (the Double-CNOT gate) that is {\em
  equally efficient} as CNOT, requiring also just three applications
together with at most eight single-qubit gates to construct any
two-qubit operation is one step in this direction~\cite{Zhang:03b}.

 We have now discovered a new quantum gate that possesses greater
 efficiency than both the CNOT and Double-CNOT gate, and that provides
 the desired minimal number of gate switchings to simulate an
 arbitrary two-qubit quantum operation.  The new gate, which we term
 the B gate, is the following:
 $B=e^{\frac\pi{2}\frac{i}2\sigma_x^1\sigma_x^2}\cdot
 e^{\frac\pi{4}\frac{i}2\sigma_y^1\sigma_y^2}$.  In the geometric
 representation illustrated in Fig.~\ref{fig:N1}, this gate is located
 at the point $B$ on the base of the tetrahedron $OA_1A_2A_3$.  This
 is a point of high symmetry, lying in the center of the region of
 gates on the base that can generate the maximum amount of
 entanglement~\cite{Zhang:02} when no local ancillas are
 allowed~\cite{Leifer:03,Kraus:01}.  This gate has the remarkable quality that
 only {\em two} applications together with at most {\em six}
 single-qubit gates can simulate any arbitrary two-qubit unitary. We
 show this by giving an analytic circuit that implements a generic
 nonlocal operation from B.  Such a generic nonlocal operation is
 $A=e^{c_1\frac{i}{2}\sigma_x^1\sigma_x^2}\cdot
 e^{c_2\frac{i}{2}\sigma_y^1\sigma_y^2}\cdot
 e^{c_3\frac{i}{2}\sigma_z^1\sigma_z^2}$~\cite{Zhang:02,Khaneja:01,Kraus:01}.
 As discussed in Ref.~\cite{Zhang:02}, the local invariants of $A$ can
 be completely defined in terms of the coefficients $[c_1, c_2, c_3]$,
 which correspond to the cartesian coordinates in Fig.~\ref{fig:N1}.
 The values of these coordinates for points in the tetrahedron
 therefore provide a complete parameterization of all possible
 nonlocal gates.

Simulation of this generic nonlocal operation can be done with the
following quantum circuit 
\vspace{0.2cm}
\setlength{\unitlength}{0.16cm}
\begin{center}
\begin{picture}(49,7)
\put(0,0){\begin{picture}(14,7)
\put(0,1.5){\line(1,0){2}}  
\put(0,5.5){\line(1,0){2}}  
\put(2,0){\thicklines\framebox(4,7)[c]{$A$}}  
\put(6,1.5){\line(1,0){2}}  
\put(6,5.5){\line(1,0){2}}
\put(8,0){\makebox(4,7)[c]{$\sim$}} 
\end{picture}}
\put(12,0){\begin{picture}(37,7)
\put(0,1.5){\line(1,0){2}}  
\put(0,5.5){\line(1,0){2}}  
\put(2,0){\thicklines\framebox(4,7)[c]{$B$}}  
\put(6,1.5){\line(1,0){2}}  
\put(6,5.5){\line(1,0){9}}
\put(15,4){\framebox(7,3)[c]{\tiny $e^{c_1\frac{i}2 \sigma_y}$}}
\put(8,0){\framebox(21,3)[c]{\tiny$e^{\beta_2\frac{i}2 \sigma_z}\cdot
e^{\beta_1\frac{i}2 \sigma_y}\cdot e^{\beta_2\frac{i}2 \sigma_z} $}}
\put(29,1.5){\line(1,0){2}}  
\put(22,5.5){\line(1,0){9}}
\put(31,0){\thicklines\framebox(4,7)[c]{$B$}}  
\put(35,1.5){\line(1,0){2}}  
\put(35,5.5){\line(1,0){2}}
\end{picture}}
\end{picture}  
\end{center}  
\vspace{0.2cm}
where the parameters $\beta_1$ and $\beta_2$ satisfy
\begin{eqnarray}
\label{eq:1}
\aligned
  \cos \beta_1 &= 1- 4 \sin^2 \frac{c_2}2 \cos^2 \frac{c_3}2,\\
\sin\beta_2&=\sqrt{\frac{\cos c_2\cos c_3}{1-2\sin^2\frac{c_2}2
    \cos^2\frac{c_3}2}} .
\endaligned
\end{eqnarray}
To prove that the above quantum circuit is indeed locally equivalent
to $A$, we follow the procedure of Refs.~\cite{Makhlin:00,Zhang:02} to
calculate the local invariants of this quantum circuit as
\begin{eqnarray}
  \label{eq:2}
\aligned
g_1&=4\cos c_1\cos^2\frac{\beta_1}2 \sin^2\beta_2,\\
g_2&=4\sin c_1 \sin \beta_1 \cos\beta_2,\\
g_3&=2(\cos^4\beta_2(\cos\beta_1+1)^2\\
&\quad+2\cos^2\beta_2(\cos^2\beta_1-2\cos\beta_1-3)\\
&\quad+4\cos^2c_1+\cos^2\beta_1+2\cos\beta_1-1).
\endaligned
\end{eqnarray}
Substituting Eq.~\eqref{eq:1} into Eq.~\eqref{eq:2}, with some
subsequent simplifications, leads to the demonstration that the local
invariants of this quantum circuit are identical to those of the
generic nonlocal operation $A$ (see Eq. (25) in Ref.~\cite{Zhang:02}).
Hence, this quantum circuit provides an analytic construction for
realization of any arbitrary two-qubit operation. Note that in this
circuit, at most six single-qubit gates are needed.

The new B gate is not a Controlled-Unitary gate. These lie on the line
$OA_1$ in Fig.~\ref{fig:N1} (note that $OL$ is equivalent to
$A_1L$~\cite{Zhang:02}). B is instead a completely new gate with
different character from the CNOT gate and any other familiar quantum
gate. It is locally equivalent to a gate that performs the operation
$|m\rangle\otimes |n\rangle \to e^{\frac\pi{4}i \sigma_x (m\oplus
  n)}|m\rangle \otimes |m\oplus n\rangle$ on the computational basis.
We can thus describe the gate B by a simple circuit in terms of
commonly used quantum gates as follows:
\vspace{0.2cm}
\setlength{\unitlength}{0.17cm}
\begin{center}  
\begin{picture}(30,7)  
\put(0,0){\begin{picture}(14,7)
\put(0,1.5){\line(1,0){2}}  
\put(0,5.5){\line(1,0){2}}  
\put(2,0){\thicklines\framebox(5,7)[c]{$B$}}  
\put(7,1.5){\line(1,0){2}}  
\put(7,5.5){\line(1,0){2}}
\put(9,0){\makebox(5,7)[c]{$\sim$}} 
\end{picture}}
\put(14,0){\begin{picture}(16,7)
\put(0,1.5){\line(1,0){16}}  
\put(0,5.5){\line(1,0){7}}  
\put(3,5.5){\line(0,-1){5}}  
\put(3,5.5){\circle*{1}}  
\put(3,1.5){\circle{2}}  
\put(7,4){\framebox(7, 3)[c]{$e^{\frac\pi{4}i \sigma_x}$}}
\put(14,5.5){\line(1,0){2}}  
\put(10.5,1.5){\circle*{1}} 
\put(10.5,1.5){\line(0,1){2.5}}
\end{picture}}  
\end{picture}  
\end{center} 
\vspace{0.2cm} 
This quantum circuit consists of a CNOT gate with the control qubit on
the top wire following a Controlled-$e^{\frac\pi{4}i \sigma_x}$ gate
with the control qubit on the bottom wire. The local equivalence
between the gate B and this quantum circuit can be proved by showing
that they both have local invariants $g_1=g_2=g_3=0$.  Given the
greater efficiency of the B gate relative to the standard CNOT for
constructing arbitrary two-qubit unitaries, it will be interesting to
explore the use of this new gate for construction of quantum compilers
and quantum algorithms~\cite{Schuch:03a,Harrow:02}, and in quantum
error correction~\cite{Gottesman:98,Steane:99}.

We now consider how the new $B$ gate may be realized in experiments.
Since the B gate is optimal in constructing a quantum circuit, this
suggests that one might always prefer to implement B as the elementary
two-qubit gate for any given physical system. However, the ease and
efficiency of constructing the elementary gates for quantum
computation from an available Hamiltonian is also a critical issue for
realization of quantum circuits~\cite{Bennett:01, Vidal:01a,
  Bremner:02}.  Optimally efficient construction of quantum operations
is realized by the quantum circuit which contains a minimal number of
single- and two-qubit gates, where {\em each} gate is itself also
implemented with a minimal application time.  From a physical
perspective, there are therefore two aspects to optimality, namely the
circuit gate count and the cost (e.g., in time) for physical
generation of individual gates from the physical Hamiltonian.  Our
result above shows that the B gate is optimal for the former.
Concerning the second question of cost in constructing gates from a
given Hamiltonian, we find that this depends on the form of the
Hamiltonian, with the B gate easier to implement than the CNOT (or
DCNOT) gate in some situations while the CNOT (or DCNOT) gate may be
easier to achieve in other situations. As a simple example, we can
consider pure nonlocal Hamiltonians, i.e., those containing no
single-qubit terms. From the geometric theory~\cite{Zhang:02}, it is
straightforward to prove that when the physical Hamiltonians are
$\sigma_z^1\sigma_z^2$ (Ising interaction),
$\sigma_x^1\sigma_x^2+\sigma_y^1\sigma_y^2$ (XY interaction), and
$2\sigma_x^1\sigma_x^2+\sigma_y^1\sigma_y^2$, it requires only one
switching to implement CNOT, DCNOT, and B gate, respectively.  The
overall circuit optimality is thus implicitly linked to the particular
choice of physical implementation.

\begin{figure}[tb]
\begin{center}
 \psfrag{A}[][]{$O$}
 \psfrag{B}[][]{$\frac\pi{4}$}
 \psfrag{C}[][]{$\frac\pi{2}$}
 \psfrag{D}[][]{$\frac{3\pi}4$}
 \psfrag{E}[][]{$\pi$}
 \psfrag{G}[][]{$\frac{\pi}4$}
 \psfrag{H}[][]{$\frac{\pi}2$}
 \psfrag{c1}[][]{$c_1$}
\psfrag{c2}[][]{$c_2$}
\psfrag{b}[][]{$B$}
\includegraphics[width=60mm]{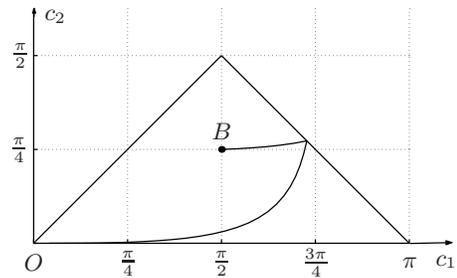}
\end{center}
\caption{Trajectory of inductively coupled Josephson junction qubits that
  generates the gate B from the Hamiltonian $H_J$.  Shown here is the
  minimum time solution, which is obtained for the Hamiltonian
  parameters $E_L=1$, $\alpha =1.1436$ with minimal time duration
  $t=1.5014$. The trajectory is confined to the $c_1, c_2$ basal plane
  of the tetrahedron $OA_1A_2A_3$ of Fig.~\ref{fig:N1}.}
\label{fig:N2}
\end{figure}

We now illustrate physical generation of the B gate with an example
where B is seen to be implemented with a shorter application time than
CNOT, and hence provides an overall optimal circuit.  We consider
charge-based Josephson qubits that are inductively
coupled~\cite{Makhlin:99}. The elementary two-qubit operations are
generated by the interaction Hamiltonian $H_J = -\frac12
E_J(\sigma_x^1+\sigma_x^2)+(E_J^2/E_L)\sigma_y^1\sigma_y^2$, where
$E_J$ is the single-qubit Josephson energy and $E_L$ is a scale
factor.  Without loss of generality we can set $E_J=\alpha E_L$, so
that $H_J=-\frac12 \alpha E_L (\sigma_x^1+\sigma_x^2)+\alpha^2
E_L\sigma_y^1\sigma_y^2$, and consider $E_L=1$.  Estimates based on
current circuit capabilities suggest that values $0.01 \leq \alpha
\simeq 1$ are feasible~\cite{Makhlin:01}.  Application of this
two-qubit Hamiltonian generates a unitary evolution $U=e^{iH_Jt}$
(with $\hbar=1$) that is characterized by the three local
invariants~\cite{Zhang:02}
  \begin{eqnarray}
    \label{eq:6}
\aligned
g_1&=\frac4{1+\alpha^2}\big(\alpha^2(x^2+y^2-1)+x^2 \big),\\
g_2&=0,\\
g_3&=\frac4{1+\alpha^2}\big(3\alpha^2-1-4y^2\alpha^2+8\alpha^2x^2y^2\\
&\qquad\qquad+4x^2-4x^2\alpha^2\big),
\endaligned
  \end{eqnarray}
  where $x=\cos{\alpha^2} t$ and $y=\cos \sqrt{(\alpha^2+1)}\alpha t$.
  The geometric theory provides us with the means to translate the
  time-dependent values of these invariants to a trajectory of points
  in the tetrahedron of Fig.~\ref{fig:N1} and hence to the specific
  nonlocal two-qubit gates that are naturally implemented as a
  function of this time evolution.  From the second invariant,
  $g_2=0$, we obtain the cartesian coordinate $c_3 = 0$ in
  Fig.~\ref{fig:N1}, which implies that all nonlocal operations that
  are generated by one application of $H_J$ must be located on the
  base of the tetrahedron $OA_1A_2A_3$.  The CNOT gate is located at
  the point $L$ in Fig.~\ref{fig:N1}, whereas the B gate is located in
  the middle of the triangular base (see also Fig.~\ref{fig:N2}).
  Each of these gates represents a different local equivalence class
  and is characterized by its own set of values $(g_1, g_2, g_3)$.
  For CNOT $(g_1, g_2, g_3)=(0,0,4)$~\cite{Zhang:02}, whereas for the
  B gate $(g_1, g_2, g_3)=(0,0,0)$. Thus to find conditions such that
  the Hamiltonian $H_J$ exactly achieves either of these gates in one
  application, we need to both tune the Hamiltonian parameter $\alpha$
  and also to find the time duration $t$ that solves Eq. (\ref{eq:6})
  for the corresponding values of $g_1$ and $g_3$. Whenever there is
  no solution, at least two switchings of the Hamiltonian will be
  required to reach the target gate.
  
  In~\cite{Zhang:02}, we have shown that the time optimal solution to
  reach CNOT in a single application is achieved when $\alpha=1.1992$,
  with application time $t=2.7309$.  In contrast, the new B gate is
  found to be reached with a shorter single application time of $H_J$
  for its optimal solution, which has a similar value of $\alpha$.
  After some algebraic work, we find that the solutions to
  Eq.~(\ref{eq:6}) for $g_1=g_3=0$ must satisfy
\begin{eqnarray}
  \label{eq:21}
  x=\cos\dfrac{2n+1}8 \pi , \quad
y^2=1-\frac{1+\alpha^2}{\alpha^2}x^2,
\end{eqnarray}
where $n$ is an integer. Hence the time $t$ at which $B$ is achieved
is $t=(2n+1)\pi/{8\alpha^2}$, where $\alpha$ satisfies
\begin{eqnarray}
  \label{eq:41}
  \gathered
\sin^2{\sqrt{1+\alpha^{-2}}\ \frac{2n+1}8
  \pi}=\frac{1+\alpha^2}{\alpha^2}\cos^2 \frac{2n+1}8 \pi.
\endgathered
\end{eqnarray}
Numerical solution indicates that the allowable values of $\alpha$
constitute an infinite set. The minimum time solution is obtained for
$\alpha=1.1436$, with corresponding minimum application time
$t=1.5014$.  The trajectory representing this time evolution through
the nonlocal gate space to the target gate B is shown in
Fig.~\ref{fig:N2}.  Thus, in this physical system the dual savings of
shorter application time of the B gate over CNOT will add to the
smaller number of gate applications required to implement an arbitrary
operation, resulting in significantly less introduction of decoherence
in experimental implementations of quantum logic and simulations that
are based on the B gate.

In summary, we have presented a new quantum gate B that provides a
direct analytic recipe to efficiently implement any arbitrary
two-qubit quantum operation with just two applications. This is more
efficient than the standard CNOT gate and Double-CNOT gate, both of
which require three applications to realize an arbitrary two-qubit
unitary.  The B gate indeed achieves the minimum number possible, as
is easily seen by recognizing that a circuit consisting of just one
application of a given two-qubit gate together with local unitaries
can only produce quantum operations that are locally equivalent to
that given gate. To simulate an arbitrary two-qubit operation with
this B gate, we only need to determine two parameters in a simple
quantum circuit. An explicit expression for these two parameters is
given here. Taken together with at most six single-qubit gates
accompanying this minimal two-qubit gate count, this provides a new
paradigm for optimally efficient construction of quantum operations.

We have illustrated the physical generation of B with the example of
inductively coupled Josephson junction qubits, for which generation of
B is also seen to be more efficient than generation of CNOT in that it
requires less application time.  In this situation, the new gate
provides an optimal route from the Hamiltonian to any arbitrary
two-qubit quantum operation, thereby providing an efficient
realization of one of the basic requirements for quantum circuit
construction.  Similar analysis can be made for other physical
implementations using the time evolution approach described above and
in Ref.~\cite{Zhang:02}.  Given these advantages of the new B gate, we
expect it will be very useful for the further development of quantum
computation.

\begin{acknowledgments}
  We thank the NSF for financial support under ITR Grant No.
  EIA-0205641. The effort of JZ, JV, and KBW is also sponsored by the
  Defense Advanced Research Projects Agency (DARPA) and the Air Force
  Laboratory, Air Force Material Command, USAF, under Contract No.
  F30602-01-2-0524.
\end{acknowledgments}

\bibliographystyle{apsrev} 

\end{document}